\documentclass[twocolumn,showpacs,showkeys,preprintnumbers,amsmath,amssymb]{revtex4}
\usepackage{docs}%
\usepackage{bm}%
\usepackage[dvips]{graphicx}
\expandafter\ifx\csname package@font\endcsname\relax\else
 \expandafter\expandafter
 \expandafter\usepackage
 \expandafter\expandafter
 \expandafter{\csname package@font\endcsname}%
\fi

\begin{document}

\title{The stochastic resonance mechanism in the Aerosol Index dynamics}%

\author{S. De Martino}%
\email{demartino@sa.infn.it} \affiliation{Dipartimento di Fisica
-Universit\'a degli Studi di Salerno, Via S. Allende, Baronissi
(SA), I-84084 Italy}

\author{M. Falanga}%
\email{rosfal@sa.infn.it} \affiliation{Dipartimento di Fisica
-Universit\'a degli Studi di Salerno, Via S. Allende, Baronissi
(SA), I-84084 Italy}

\author{L. Mona}%
\email{mona@imaaa.pz.cnr.it} \affiliation{Dipartimento di Fisica
-Universit\'a degli Studi di Salerno, Via S. Allende, Baronissi
(SA), I-84084 Italy\\ Dipartimento di Ingegneria e Fisica
dell'Ambiente - Universit\'a della Basilicata,C.da Macchia Romana,
Potenza (PZ),I-85100 Italy}

\date{January 2002}%



\
\
\begin{abstract}
We consider Aerosol Index (AI) time-series extracted from TOMS
archive for an area covering Italy  $(7-18^o E ; 36-47^o N)$. The
missing of convergence in estimating the embedding dimension of
the system and the inability of the Independent Component Analysis
(ICA) in separating the fluctuations from deterministic component
of the signals are evidences of an intrinsic link between the
periodic behavior of AI and its fluctuations.  We prove that these
time series are well described by a stochastic dynamical model.
Moreover, the principal peak in the power spectrum of these
signals can be explained whereby a stochastic resonance, linking
variable external factors, such as Sun-Earth radiation budget and
local insolation, and fluctuations on smaller spatial and temporal
scale due to internal weather and antrophic components.
\end{abstract}
\maketitle

\section{Introduction}
\label{sec:intro}
 Over the last two decades, much attention has been
devoted to study stochastic resonance (SR) as a model for many
kinds of physical phenomena. The term "stochastic resonance"
describes a phenomenon whereby a weak signal can be amplified and
optimized by the presence of noise. Stochastic resonance was
introduced by  Benzi et al. \cite{benzi,benzi1983} and Nicolis
\cite{nicolis} in the field of the physics of atmosphere to
explain the periodic recurrent ice ages. They used the two levels
Budyko-Sellers potential \cite{Budyko} for describing the incoming
and outcoming radiation and added to it a weak periodic forcing,
representing the small modulation of the Earth's orbital
eccentricity and a noise corresponding to variations on small
time-scale respect to eccentricity. They found that the presence
of the noise amplifies the periodic component, facilitating the
jump of the system between the two levels synchronously with the
periodical forcing. So they obtained ice ages as a peak, in the
power spectrum of a temperature record, centered around the
$100000$ year period and characterized by the correct amplitude.
Since then, stochastic resonance has been observed in a large
variety of physical systems, including bistable ring lasers,
semiconductor devices, chemical reactions and neurophysiological
systems \cite{gammaitoni}. In the atmospheric field, only recently
it appears another application of the SR. In fact, in 1994
El-Ni\~{n}o-Southern-Oscillation has been modeled with stochastic
resonance with a periodic forcing of five years
\cite{Tziperman,Stone} i.e. stochastic resonance is conjectured to
have some relevance also at local climate scale. From these works,
however, it seems to be a strong link between the fluctuations in
the atmospheric medium (humidity, temperature, wind speed and
direction, industrial emission and urban pollution) and the
dynamics, that describes the atmospheric behavior on larger
spatial and temporal scales. In this paper, we  want to reconsider
the relevance of SR on global climate scale but on smaller time
scale, considering that periodic behaviours of Earth motion could
induce stochastic resonance. In this study, we consider
time-series representing distribution of tropospheric aerosols
that surely are affected by Earth's revolution and rotation
motions. Actually much attention is devoted to a better
understanding of the aerosol role in the radiation budget
\cite{Andreae}. The aerosol effect on climate is quantified in
terms of radiative forcing, namely the net variation of radiation
flux at the top of the atmosphere due only to radiative aerosol
effects.  A cooling effect of some aerosol (ash and sulfate
aerosols) is well-documented \cite{Charlson1987}, while other
types absorb solar radiation (dust, carbonic and silicaceous
aerosols). At present, there are many activities addressed to
study the optical properties of anthropogenic aerosols, for
determining their influence on the radiative forcing
\cite{Tegen1996,Charlson}. The finding of SR into time series from
troposheric aerosol should give some new insight on this topic.

\section{Datasets}
\label{sec:data} We extracted time-series to analyze from the
archive of TOMS (Total Ozone Mapping Spectrometer) Aerosol Index
(AI hereafter). In particular we use the data retrieved from the
measurements effected by Nimbus 7, an eliosynchronous satellite
orbiting around the Earth from November 1978 to May 1993. The AI
was determined using the backscattered radiances $(I_{340} ,
I_{380})$ measured by TOMS at 340 and 380nm in the following way:

\begin{eqnarray}
\label{AI} \nonumber
AI=-100\left[log\left(\frac{I_{340}}{I_{380}}\right)_{meas}
-log\left(\frac{I_{340}}{I_{380}}\right)_{theor}\right]
\end{eqnarray}

where the subscripts \textit{meas} and \textit{theor} indicate
respectively measured and theoretically computed quantities
\cite{Torres1998}. This parameter gives an information on the
aerosol optical depth and strongly depends  on aerosol layer
height and on the optical properties of the aerosols
\cite{Hsu1999}. The AI is defined in such a way that it is
positive for UV absorbing aerosols and negative for no absorbing
ones, even if it assumes small negative values when absorbing
aerosols are near the Earth's surface, up to 1.5 km
\cite{Hsu1999}. It is important to underline that the clouds are
characterized by a null AI, but their influence on the
measurements is not negligible. In fact the presence of clouds can
act as a screen for the detection operations, at least in the case
of clouds covering large areas \cite{Torres1998}. The data have a
spatial resolution of $1^o$ latitude x $1.25^o$ longitude
(corresponding to 50km x 50km) and temporal resolution of one
sample a day. At this time, we consider only a restricted area
relative to Italy with the following geographical coordinates
$(7-18^o E ; 36-47^o N)$. To reduce the clouds contribution on our
data, we used the reflectivity data from TOMS as parameter to
distinguish the case of obscuring clouds from not one. If the
reflectivity exceeds a threshold of $25 \%$, we consider the AI as
corrupted by clouds and we corrected this sample replacing the
experimental value with the previous one plus a quantity chosen
randomly among $0$, $0.1$ and $-0.1$, where $0.1$ is the
instrumental error. The physical hypothesis underlying this
correction is the small variability of the aerosol content in two
days on a area of 50km x 50km. Moreover we have to take into
account that the Nimbus 7 stability period started from the 1984,
so we consider time-series extended from January 1984 up to May
1993 \cite{Herman}.

\section{Analysis}
The first step in examining AI data is to use standard linear and
nonlinear analysis techniques to prove that we are in presence of
a dynamics intrinsically stochastic. In fig. 1 we can see one
scalar time-series and its power spectrum. The power spectrum
shows two evident peaks: the principal peak corresponds to an
annual period, while the second one is related to a period of
about 6 days.
\begin{figure}
\includegraphics[width=7cm]{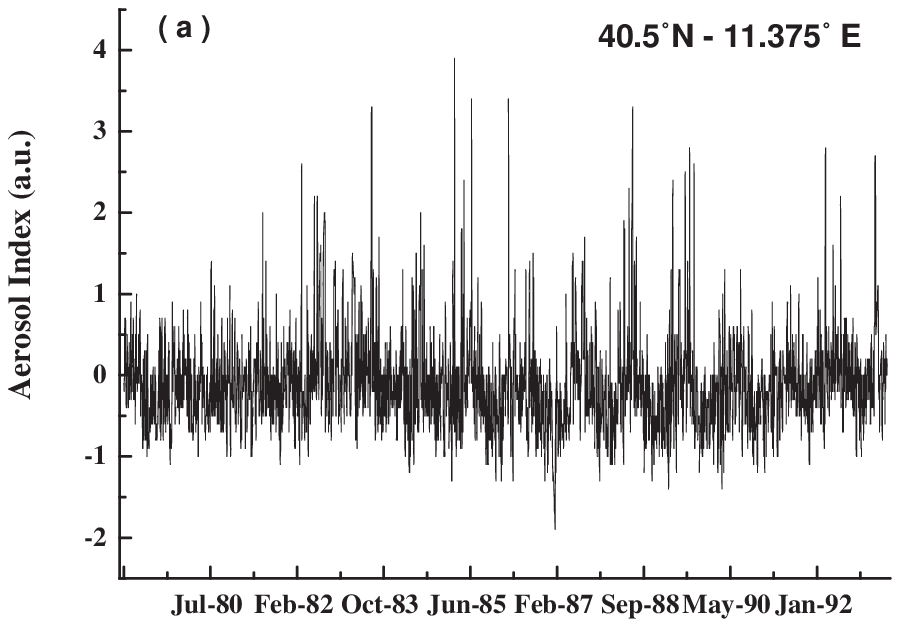} 
\end{figure}
\begin{figure}
\includegraphics[width=7cm]{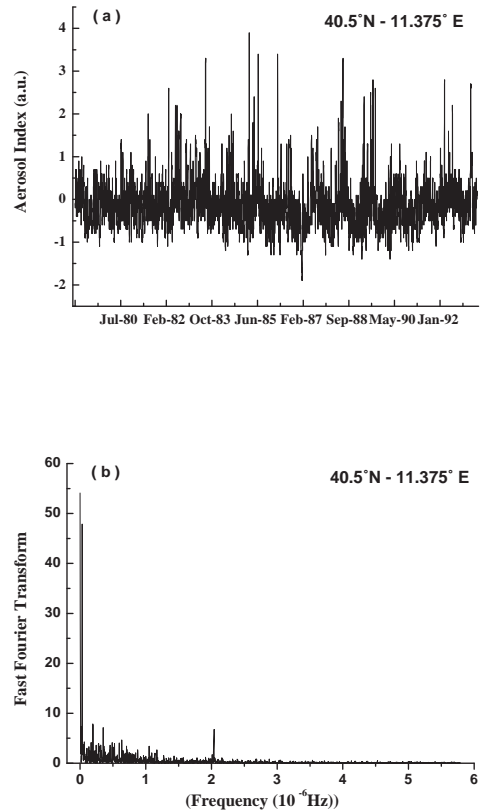}
\caption{\label{fig:exsign} Example of Aerosol Index time-series
(a) and its Power Spectrum (b) after the correction for
reflectivity. Both the parts of the figure are related to an area
of $1^o$ latitude x $1.25^o$ longitude centered at $40.5^o$ N -
$11.375^o$ E.}
\end{figure}
For upgrading our knowledge about the dynamics generating these
signals, as a first step, we apply the Independent Component
Analysis (ICA) (\cite{Hyvarinen1999} and reference therein). It is
a well established method, based on Information Theory, to
extract, from a recorded signal, independent dynamical systems
also non linear, but linearly superposed \cite{Acernese2000}. The
ICA separates also noise.
\begin{figure}
\includegraphics[width=7cm]{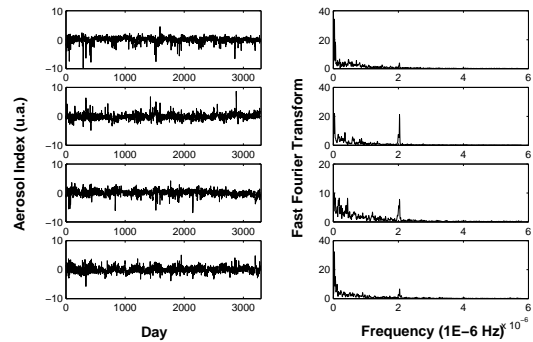}
\caption{\label{fig:ica} Result of the ICA application on 99
time-series recorded over Italy. We obtain 4 signals, but each of
them still contains memory of the annual and weekly peaks as well
as of the noise.}
\end{figure}
As we can observe looking at Fig.2, the application of ICA to our
signals has not been able to separate noise contribution from
deterministic one. So we may conclude that the periodic part of
the signal is intrinsically tied to the fluctuations, i.e. our
signals are generated by a genuine stochastic dynamics. Another
confirmation of the stochastic nature of the dynamics can be
derived from estimating embedding dimension. This estimate is made
applying the False Nearest Neighbors method \cite{Kennel1992} to a
single record: we obtain that up to dimension equal to 10 the
algorithm does not converge as in the case of a stochastic
process. In conclusion analyzing time series, it seems to be
plausible to conjecture the presence of a mechanism of SR acting
between the stochastic noise and the annual periodicity. It is
important stressing that, if we conjecture SR as a global
mechanism, it is also true that the parameters that rules the SR
(local radiation budget and insolation) are characteristic of the
site location. So it is clear that these parameters, as well as
the noise level, depend on the width of the investigated area. In
this first step, we focalize our attention on Italy $(7-18^o E ;
36-47^o N)$: we choose this particular area for investigating the
contributes of both antrophic and natural aerosol sources and
their mutual influence. Thus we have extracted 99 recorded series
considering each one of them as a realization of the same
stochastic process in a different spatial point. Since a
comparison between two stochastic process can be made only on
average, we have to extract the average series of experimental
data that will be compare with that derived from the model. In
this first analysis we devote our attention only to the dominant
annual peak, so we execute a pass-band linear filter in the range
of $10^{-9}-10^{-6}$ Hz. Successively, we eliminate from the
signal a trend that can be ascribed to a residual trace of the
satellite instability, obtaining the time-series represented in
Fig.3.
 Now, finally, we have the experimental average
time-series to be compared with the one derived from a simulated
model.

\section{Modelling}
\begin{figure}
\includegraphics[width=6cm]{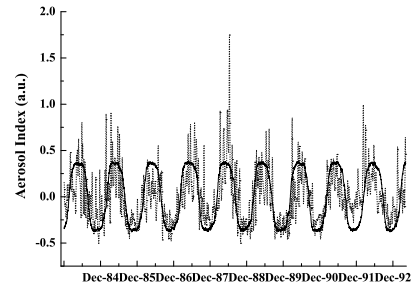} 
\caption{\label{fig:confr} Experimental average signal relative to
Italy $(7-18^o E ; 36-47^o N)$, filtered in the $10^{-9}-10^{-6}$
Hz range and subtracted of a residual trend [dotted line] and
simulated one obtained averaging on 100 signals [solid line]. The
parameters used in this simulation are $\textit{a}=0.3636$,
$\textit{$A_0$}=437*10^{-4}$ and $ \nu=0.06$ }.
\end{figure}
\begin{figure}
\includegraphics[width=6cm]{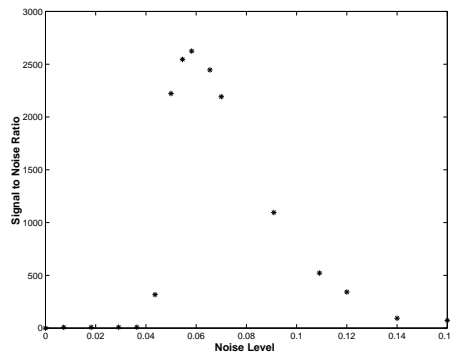}
\caption{\label{fig:snr} Estimate of the Signal to Noise Ratio
 obtained from the signal averaged over
100 realizations for each value of the noise level.}
\end{figure}
 We have to mark two goals, the first is to numerically
prove that our physical system can be effectively described with a
resonance model, the second one is to make the best fit of the
involved parameters and to give a physical interpretation of them.
In the framework of the model used by Benzi et al.
\cite{benzi,benzi1983}, we described our physical system by means
of the following equation:

\begin{eqnarray}
\label{dx} \nonumber
dx=\left[\left(x-\frac{x^3}{a}\right)+Acos(\omega t)\right]dt+\nu
\end{eqnarray}

So we have considered a double well potential representing the
radiative balance of incoming and outcoming radiation at the top
of the atmosphere, that should be directly correlated with the
different height of the aerosol layers in winter and summertime.
This potential is completely characterized by the gap between them
(gap that we indicate with \textit{a}). The periodical component
has a frequency of $3.2*10^-8$ Hz, corresponding to the annual
peak, and amplitude \textit{$A$}. This term is included to take
into account the change of daily insolation with the solar
declination angle. Because the short period disturbances can be
described as a random walk \cite{hasselmann}, the noise has been
included as a Wiener process multiplied for a diffusion
coefficient $\nu$. The estimate of parameters will give us the
noise level, the amplitude of periodic forcing and then the
estimate of difference between the two levels. Bearing in mind
that, in regime of SR, the oscillation amplitude is equal to
\textit{a}, we fix this parameter in agree with the experimental
average at the value of $0.3636$. So we have reduced the problem
to determine the value of two parameters
 to better approximate our experimental signal.

Tuning the parameters we realize the time scale conditions
\cite{gammaitoni} finding the values that give us a maximum
correlation integral between the mean value of the simulated
stochastic process (obtained from 100 realizations for each value
of the parameters) and the averaged experimental signal. In this
way we find a maximum correlation integral of 0.71 corresponding
to a
 dimensionless level noise of about 0.06 and an amplitude of periodic forcing equals to $437*10^{-4}$.
The last quantity corresponds to the 10 percent of insolation
variation over Italy, respect to the clear day solar insolation
(estimated as $1000W/m^2$). With this simulation, we have the
estimation of the parameters describing our system and we can also
reconstruct the Signal to Noise Ratio (SNR) behavior as function
of the noise strength. The SNR curve represented in Fig.4 is
characteristic of the stochastic resonance \cite{gammaitoni}: this
confirms us that the AI dynamics can be described with a SR
mechanism. Moreover the best approximating signal corresponds to a
noise level of $\nu$=0.06, so in this curve of resonance we are
around the maximum.

\section{Conclusions}
The stochastic nature of the atmospheric medium is a
well-established reality,  but the enhancing function of this
component is a no-trivial one. In this paper, we have proved that
the Aerosol Index dynamics over Italy can be described through a
stochastic resonance model and in addition we have found that we
are effectively in stochastic resonance regime. This is an
important aim, because it is the first time that SR is observed in
atmospheric physics on human time scale. Surely this could be a
fundamental tool for a better understanding of atmospheric
dynamics, in particular for what concerns the links between the
mechanism of global circulation and the local processes of
aerosolic production and distribution. Moreover the presence of an
atmospheric SR mechanism on a small time scale for the Aerosol
Index has certainly some direct and indirect effects on other
relevant atmospheric parameters such as the water vapor and
precipitable water atmospheric contents, because of the aerosol's
role as condensation nuclei.

\end{document}